\documentclass[conference]{IEEEtran}
\IEEEoverridecommandlockouts
\usepackage{cite}
\usepackage{amsmath,amssymb,amsfonts}
\usepackage{algorithmic}
\usepackage{graphicx}
\usepackage{textcomp}
\usepackage{array}
\usepackage[caption=false]{subfig}
\usepackage{booktabs}
\usepackage{pifont}
\usepackage{tabularx}
\usepackage[table]{xcolor}
\usepackage{enumitem}
\usepackage{url}
\usepackage{todonotes}
\usepackage{framed}
\usepackage{multirow}
\usepackage{pifont}
\usepackage{etoolbox}
\usepackage[many]{tcolorbox}
\usepackage{wrapfig,lipsum}
\usepackage[export]{adjustbox}
\usepackage{siunitx}
\usepackage{makecell}
\usepackage{soul}
\usepackage[bookmarks=false]{hyperref}

\usepackage{caption}
\usepackage{array}
\newcolumntype{x}[1]{>{\centering\arraybackslash\hspace{0pt}}p{#1}}

\def\BibTeX{{\rm B\kern-.05em{\sc i\kern-.025em b}\kern-.08em
    T\kern-.1667em\lower.7ex\hbox{E}\kern-.125emX}}

\newcommand{\TODOin}[1]{\todo[size=\tiny]{Sona:}}

\graphicspath{{figures/}} %Setting the graphicspath

\newcolumntype{L}[1]{>{\raggedright\let\newline\\\arraybackslash\hspace{0pt}}m{#1}}
\newcolumntype{C}[1]{>{\centering\let\newline\\\arraybackslash\hspace{0pt}}m{#1}}
\newcolumntype{R}[1]{>{\raggedleft\let\newline\\\arraybackslash\hspace{0pt}}m{#1}}
\newcolumntype{Y}{>{\centering\arraybackslash}X}

\begin{document}

%\title{Data-driven Analysis for Design Patterns in \MD{Collective Adaptive Systems}}

\title{Learning to Learn in Collective Adaptive Systems: Mining Design Patterns for Data-driven Reasoning}

\author{\IEEEauthorblockN{
Mirko D'Angelo\IEEEauthorrefmark{1},
Sona Ghahremani\IEEEauthorrefmark{2},
Simos Gerasimou\IEEEauthorrefmark{4},\\
Johannes Grohmann\IEEEauthorrefmark{7},
Ingrid Nunes\IEEEauthorrefmark{5},
Sven Tomforde\IEEEauthorrefmark{6}, 
Evangelos Pournaras\IEEEauthorrefmark{3}}
\IEEEauthorblockA{\\\IEEEauthorrefmark{1}Linnaeus University, V\"{a}xj\"{o}, Sweden, Email: mirko.dangelo@lnu.se}  
\IEEEauthorblockA{\IEEEauthorrefmark{2}
Hasso Plattner Institute, Universit\"{a}t Potsdam, Potsdam, Germany, Email: sona.ghahremani@hpi.de} 
\IEEEauthorblockA{\IEEEauthorrefmark{4}University of York, United Kingdom, Email: simos.gerasimou@york.ac.uk}
\IEEEauthorblockA{\IEEEauthorrefmark{7}University of W\"{u}rzburg, Germany, Email: johannes.grohmann@uni-wuerzburg.de}
\IEEEauthorblockA{\IEEEauthorrefmark{5}Universidade Federal do Rio Grande do Sul, Porto Alegre, Brazil, Email: ingridnunes@inf.ufrgs.br}
\IEEEauthorblockA{\IEEEauthorrefmark{6}Christian-Albrechts-Universität zu Kiel, Germany, Email: st@informatik.uni-kiel.de}
\IEEEauthorblockA{\IEEEauthorrefmark{3}University of Leeds, United Kingdom, Email: e.pournaras@leeds.ac.uk}
}

\maketitle

\begin{abstract}
Engineering collective adaptive systems~(CAS) with learning capabilities is a challenging task due to their multi-dimensional and complex design space. Data-driven approaches for CAS design could introduce new insights enabling system engineers to manage the CAS complexity more cost-effectively at the design-phase. This paper introduces a systematic 
approach to reason about design choices and patterns of learning-based CAS. Using data from a systematic literature review, reasoning is performed with a novel application of data-driven methodologies such as clustering, multiple correspondence analysis and decision trees. The reasoning based on past experience as well as supporting novel and innovative design choices are demonstrated. 
\end{abstract}


\section{Introduction}

\noindent
Collective adaptive systems (CAS) are distributed systems comprising multiple heterogeneous agents. Each agent does not individually possess system-wide knowledge and can: $(i)$~interact with other agents either directly or indirectly; $(ii)$~exhibit learning capabilities to expand its personal knowledge; and $(iii)$~make decisions based on collective or aggregated knowledge from its peers~\cite{mitchell2005self,Pournaras2018,Dangelo:2020}. 

Employing agents with such characteristics allows constructing highly autonomous systems exhibiting self-adaptive properties.
As a result, learning-based CAS can cope with uncertainties and adapt accordingly to fulfill their requirements and improve their performance and reliability.\footnote{\url{http://www.focas.eu/manifesto/}}

Engineering learning-based CAS is complex due to their non-deterministic and highly dynamic operational environment, emerging from simultaneous interactions of several autonomous entities. Moreover, system-wide knowledge is distributed among the agents, entailing that advanced mechanisms should be used for efficient knowledge acquisition and sharing. Finally, the use of learning adds yet another layer of complexity, influenced by the availability of data, choice of technique, model instantiation, and model update. Thus, the numerous influencing decisions emerging from such a multi-dimensional and the complex design space perplex the engineers' choices when designing learning-based CAS.

With the growing availability of software engineering data, data-driven techniques have emerged as an effective methodology to provide software practitioners with up-to-date and pertinent information supporting the decision-making process~\cite{Xie:2009}. Data-driven methodologies support dimensionality reduction by recognizing correlations in data~\cite{Robillard:2010} and have been extensively applied in the literature, e.g., to understand software evolution~\cite{Ball:97} or to discover instances of design patterns from the system's source code~\cite{dong:2009}.

Collecting data about relevant past experiences, extracting knowledge from it, and making the knowledge available in a manner that can be reasoned upon is a first step towards supporting CAS engineers in navigating through the large design space of such systems and making cost-effective choices~\cite{Larose2006DataMM}.

This paper contributes to this direction in the following ways. $(i)$~We extend our previous systematic literature review on learning-enabled CAS~\cite{DAngelo:2019} and use the analysis results as data capturing relevant past experiences.
$(ii)$~By employing data-driven methodologies, we identify correlations in the collected data, based on which we present relevant design-time reasoning knowledge in the form of design guidelines. And $(iii)$~we structure the data (i.e.,~past experiences) as a decision tree representing a reasoning knowledge that can serve either as a design-time recommender or to spot design gaps.

The paper is organized as follows. Section~\ref{sec:methodology} introduces the data acquisition process and the data-driven methodologies. Section~\ref{sec:application} presents the design guidelines elicited from the analysis and shows the knowledge as a decision tree. Section~\ref{sec:conclusion} concludes the findings and discusses future work.

\begin{figure*}[hbtp]
\vspace{-0.45cm}
   \subfloat[Silhouette values]{\includegraphics[width=0.155\textwidth,valign=b]{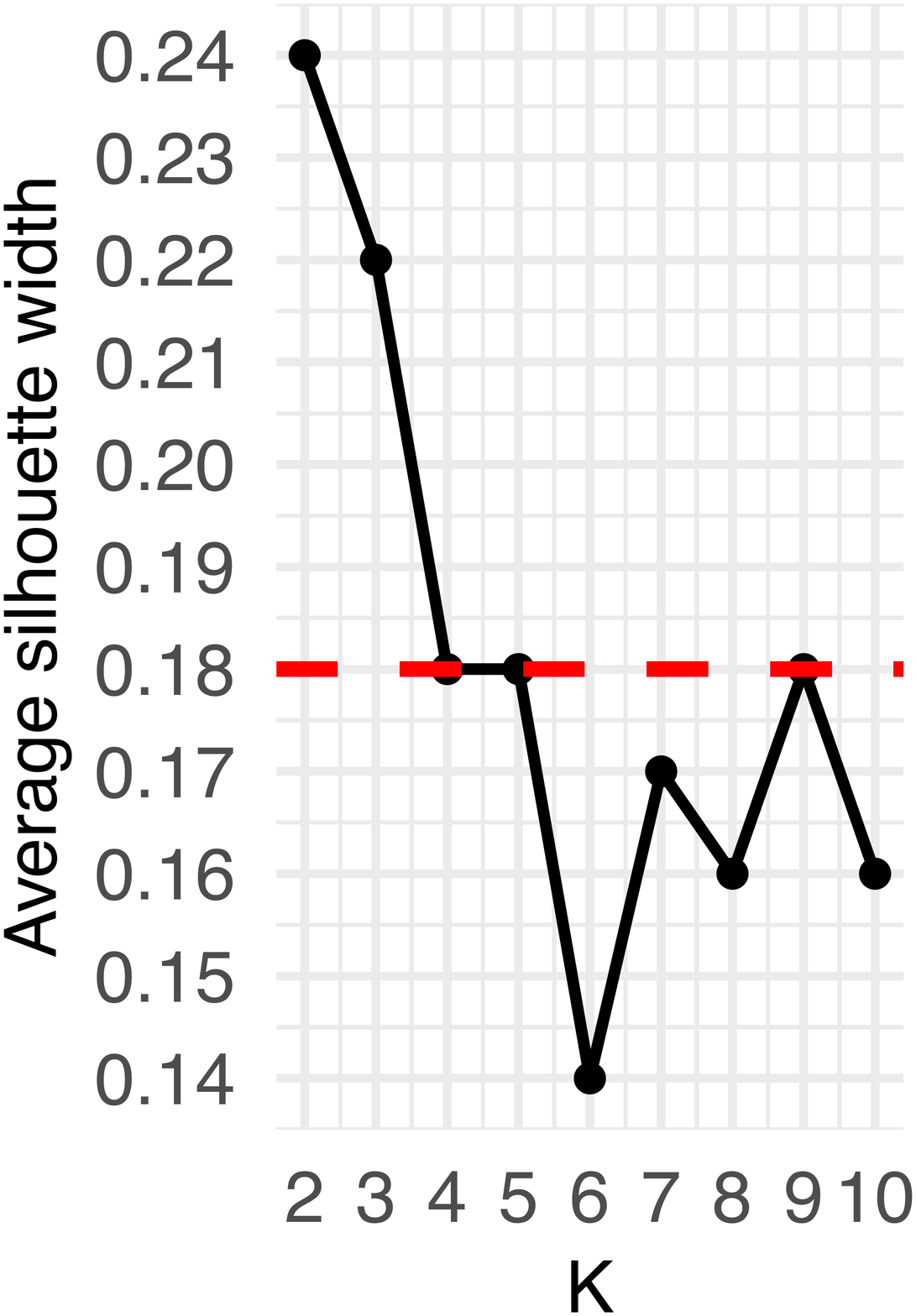}\label{fig:sil}}
    \hfill
   \subfloat[][Bootstrapping results]{
   \scriptsize
\begin{tabular}[b]{l p{4.2cm} >{\raggedleft\arraybackslash}p{3.2cm}}
\toprule
\textbf{K} & \textbf{Vector of clusters stabilities} & \textbf{Times each cluster is dissolved in 100 re-sampling}\\ \midrule
2 & $\langle$0.90,0.85$\rangle$ & $\langle$0,0$\rangle$ \\ \midrule
3 & $\langle$0.82,0.86,0.50$\rangle$ & $\langle$2,0,59$\rangle$ \\ \midrule
4 & $\langle$0.87,0.86,0.60,0.61$\rangle$ & $\langle$0,1,48,9$\rangle$ \\ \midrule
5 & $\langle$0.84,0.77,0.64,0.69, 0.27$\rangle$ & $\langle$0,8,43,29,92$\rangle$ \\ \midrule
9 & $\langle$0.70,0.69,0.61,0.64,0.84,0.74,0.47,0.51,0.33$\rangle$ & $\langle$18,32,40,36,21,29,67,69,93$\rangle$ \\ \bottomrule
\label{tab:bootstrap}
\end{tabular}
}
  \hfill
  \subfloat[$K=2$]{\includegraphics[width=0.224\textwidth]{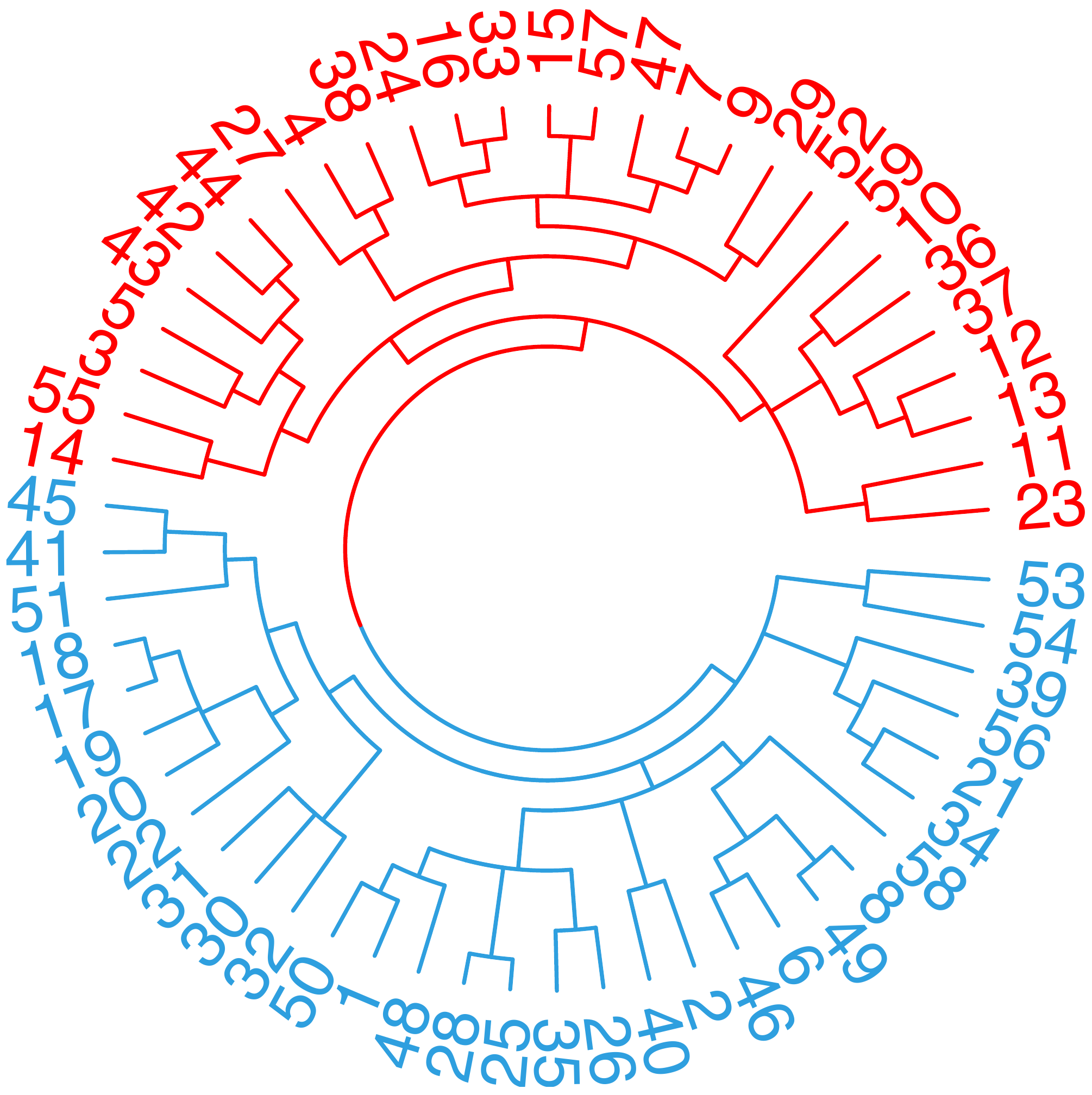}\label{fig:2cl}}
  \vspace{-0.45cm}
  \subfloat[$K=3$]{\includegraphics[width=0.224\textwidth]{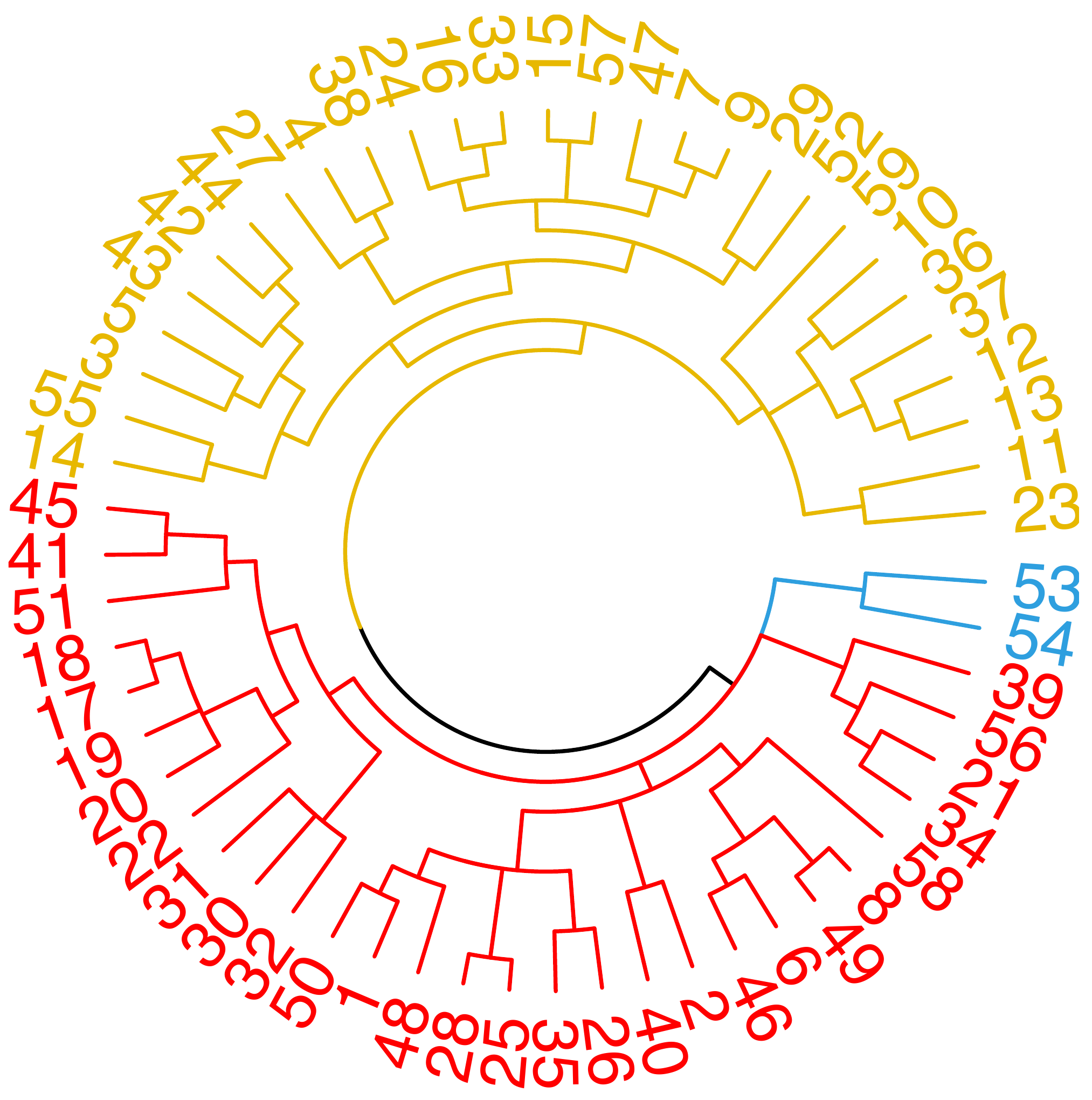}\label{fig:3cl}}
  \hfill
  \subfloat[$K=4$]{\includegraphics[width=0.224\textwidth]{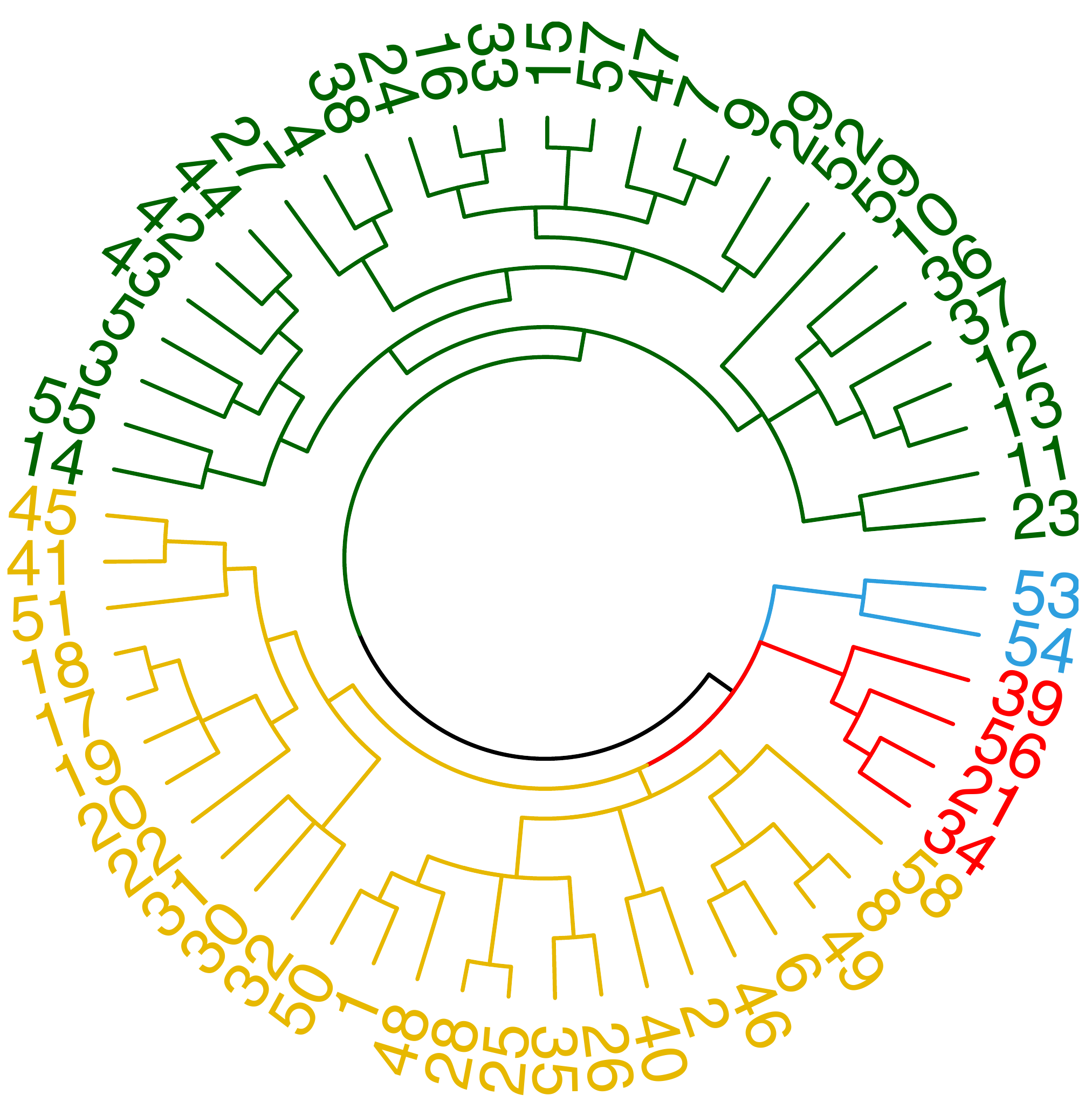}\label{fig:4cl}}
  \hfill
  \subfloat[$K=9$]{\includegraphics[width=0.224\textwidth]{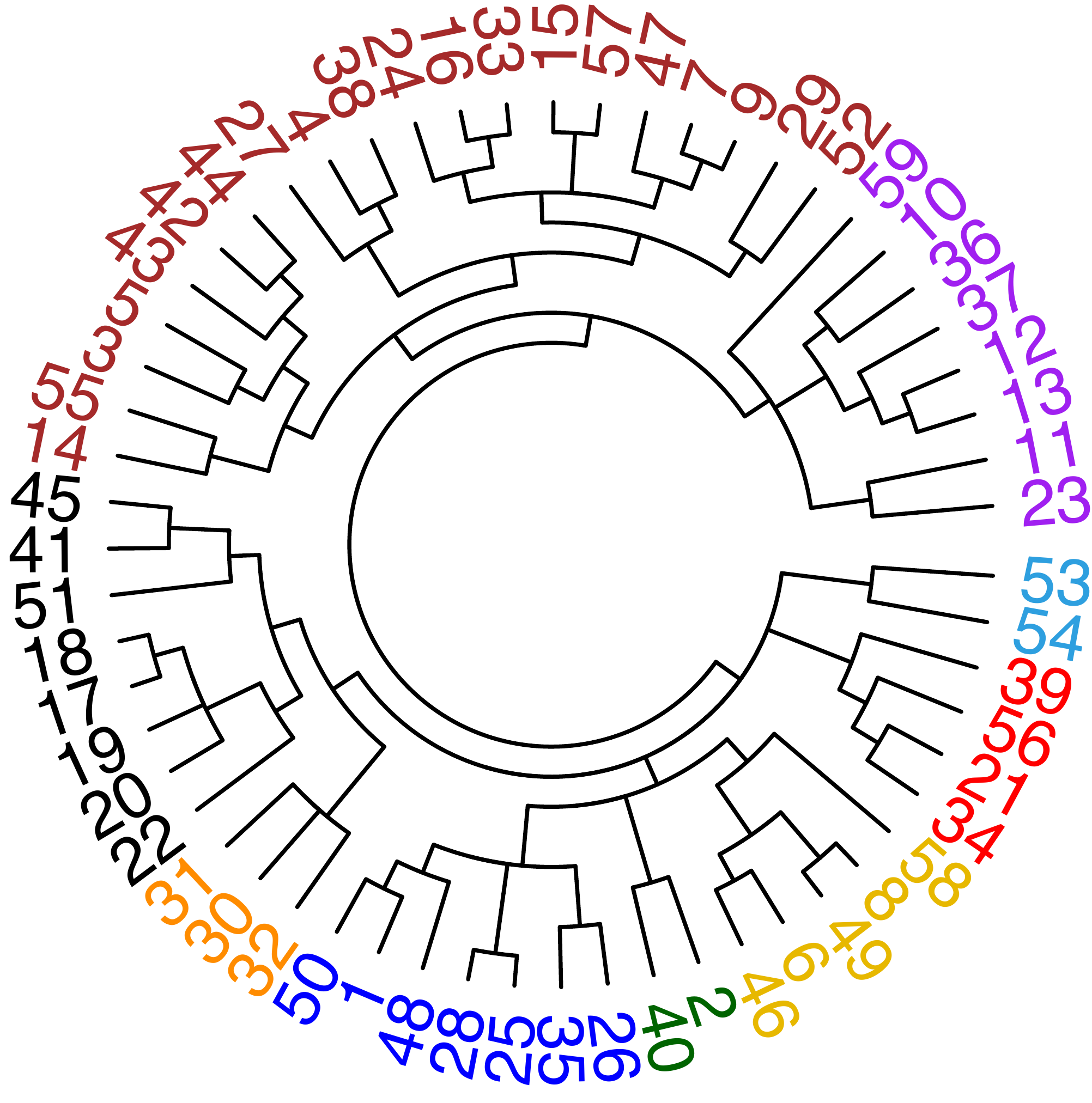}\label{fig:9cl}}
  \hfill
  \subfloat[Autonomy]{\includegraphics[width=0.224\textwidth,valign=b]{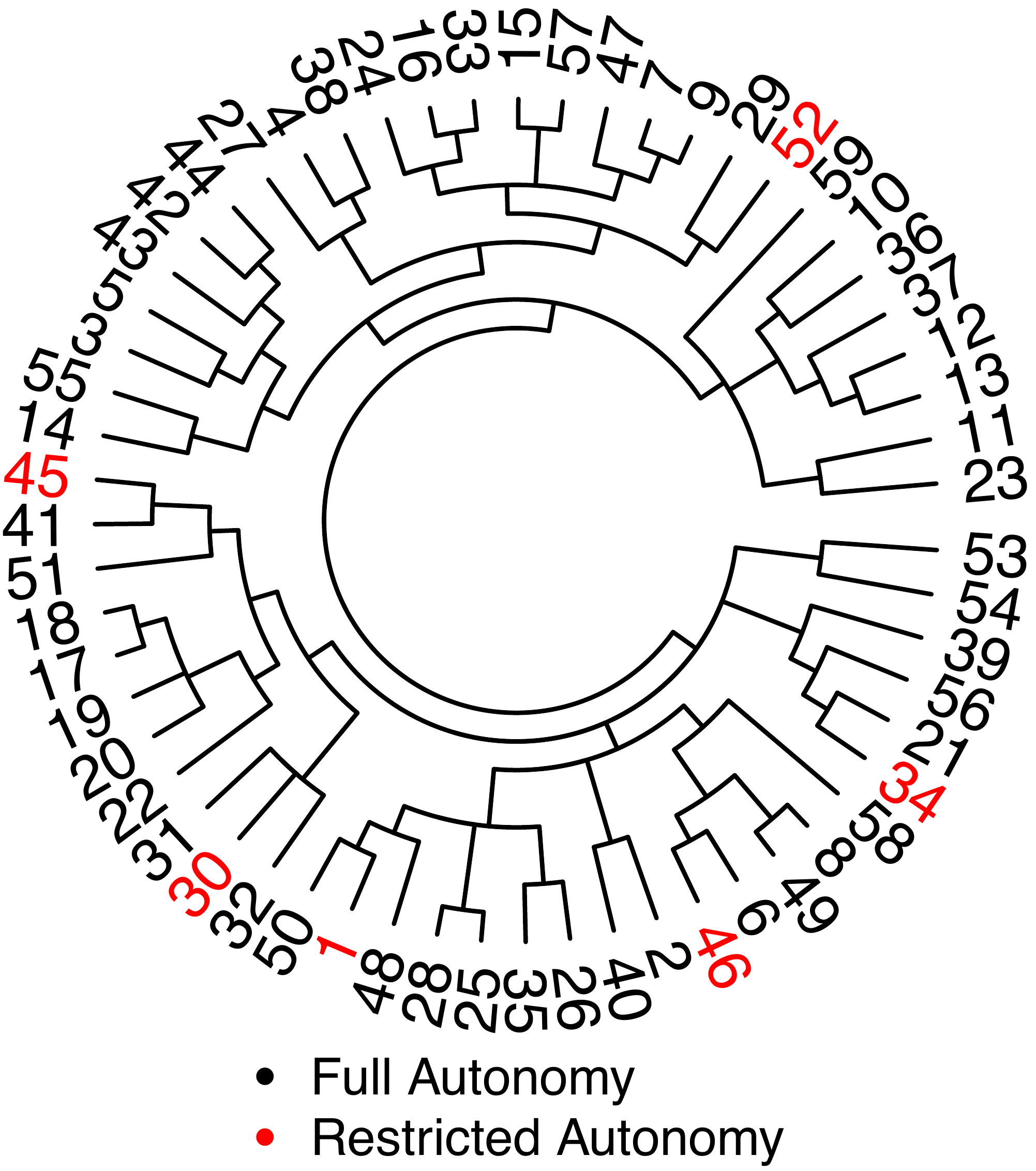}\label{fig:autonomy-ind}}
   \vspace{-0.37cm}
  \subfloat[Emergent behaviour]{\includegraphics[width=0.224\textwidth,valign=t]{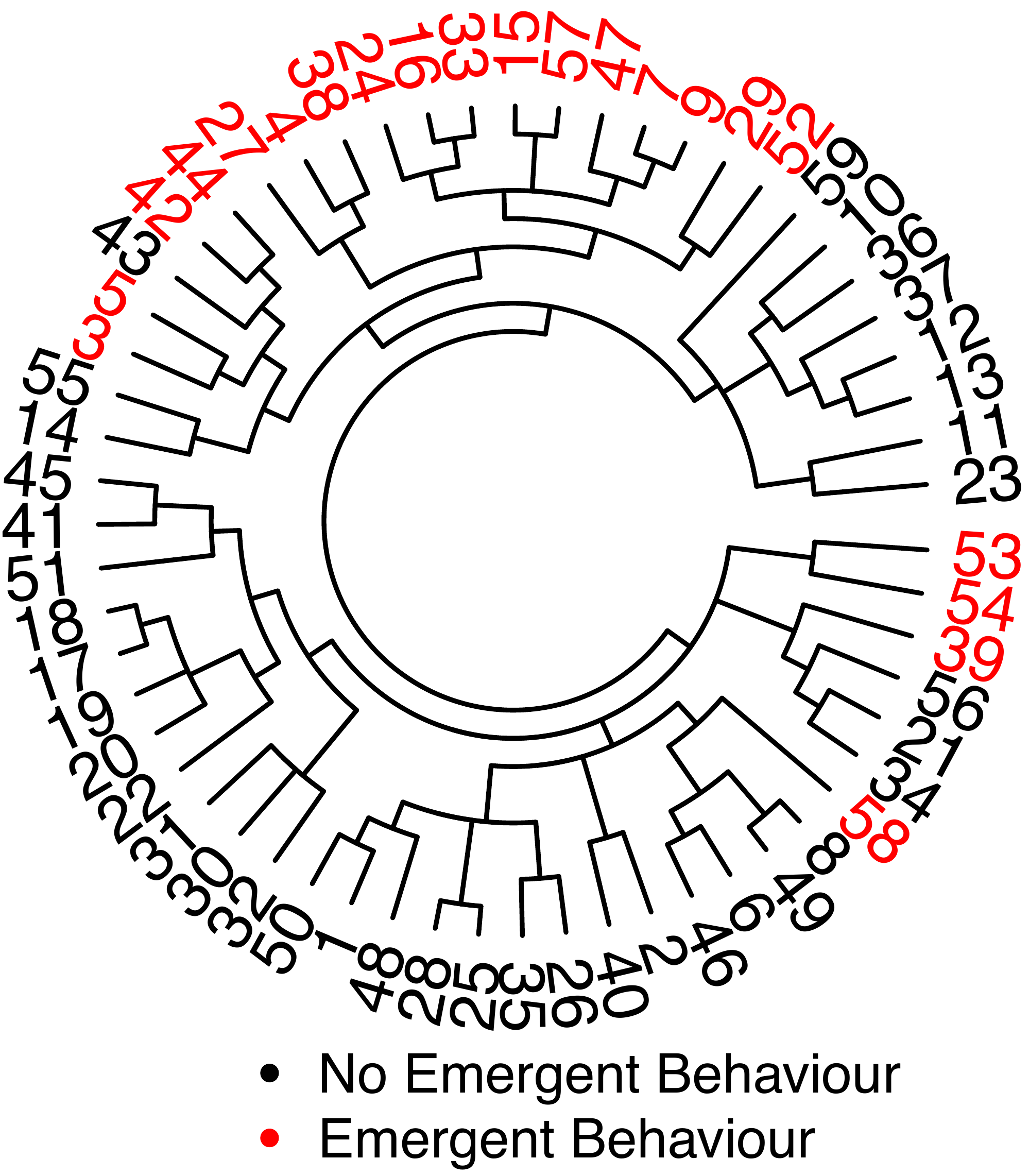}\label{fig:eb-ind}}
 \hfill
  \subfloat[Cooperative agent]{\includegraphics[width=0.224\textwidth,valign=t]{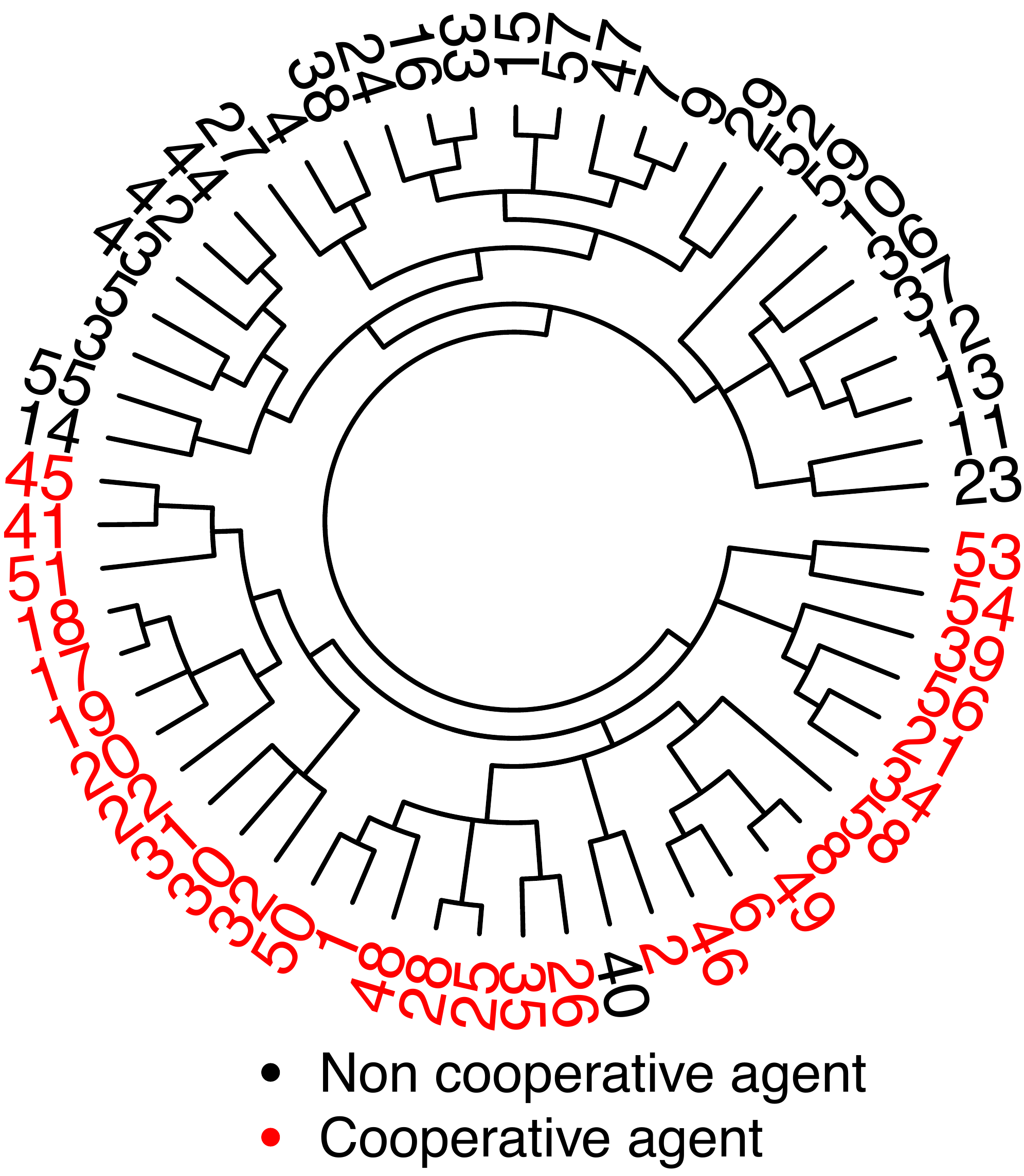}\label{fig:coop-ind}}
  \hfill
    \subfloat[Trigger first]{\includegraphics[width=0.224\textwidth,valign=t]{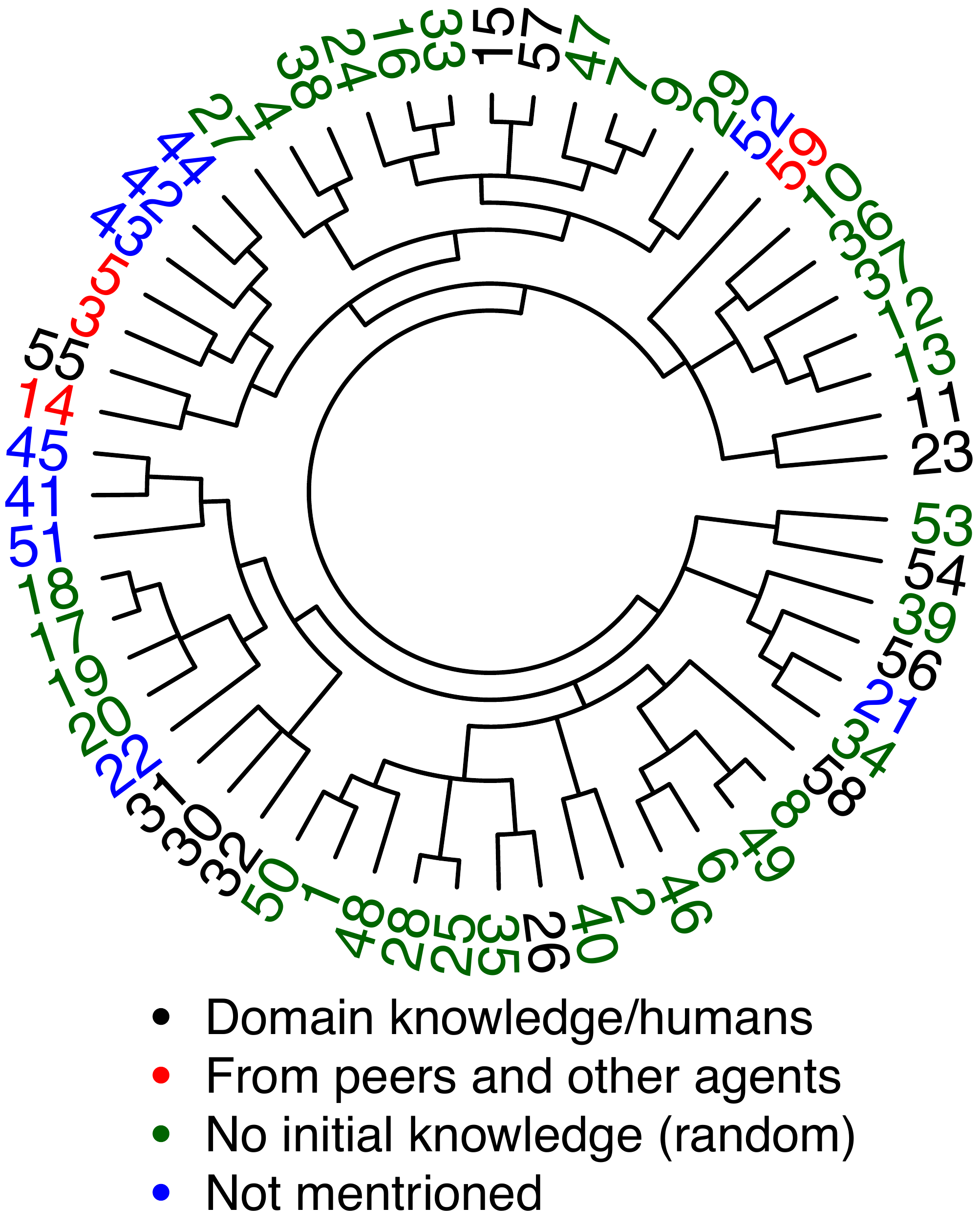}\label{fig:first-ind}}
    \hfill
  \subfloat[Trigger update]{\includegraphics[width=0.224\textwidth,valign=t]{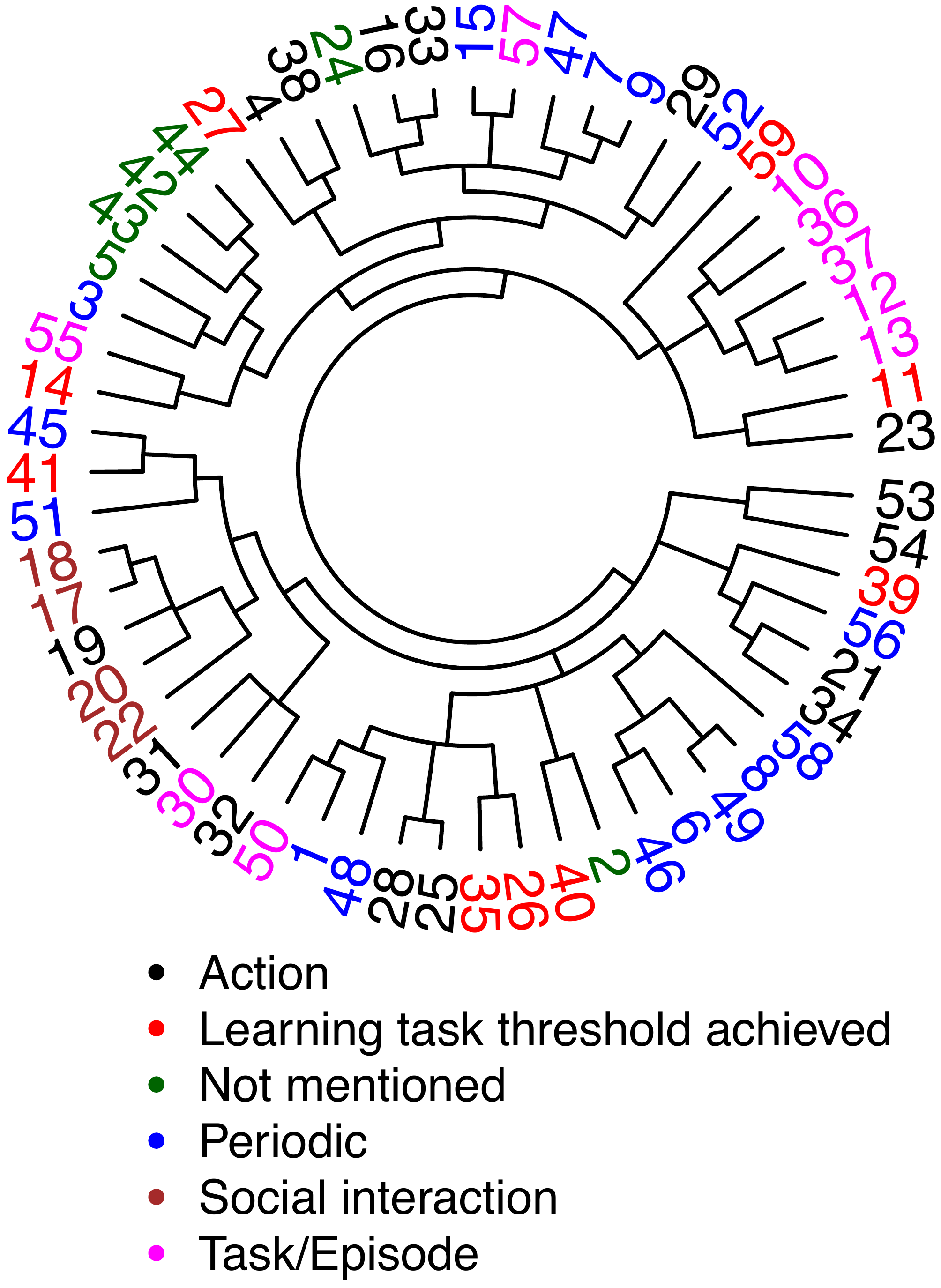}\label{fig:update-ind}}
   \vspace{-0.35cm}
  \subfloat[Behaviour]{\includegraphics[width=0.224\textwidth,valign=t]{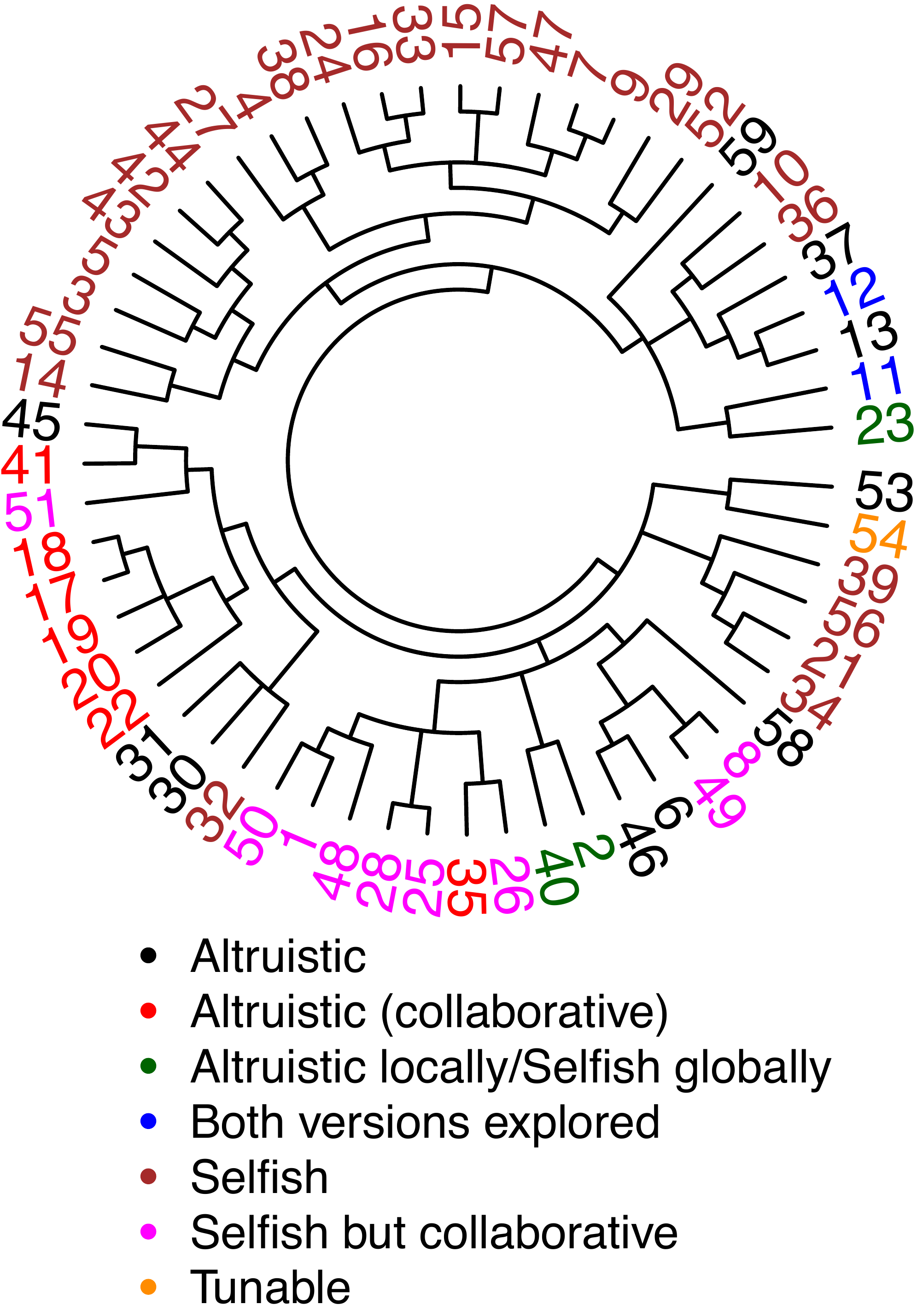}\label{fig:self-ind}}
  \hfill
  \subfloat[Knowledge access]{\includegraphics[width=0.224\textwidth,valign=t]{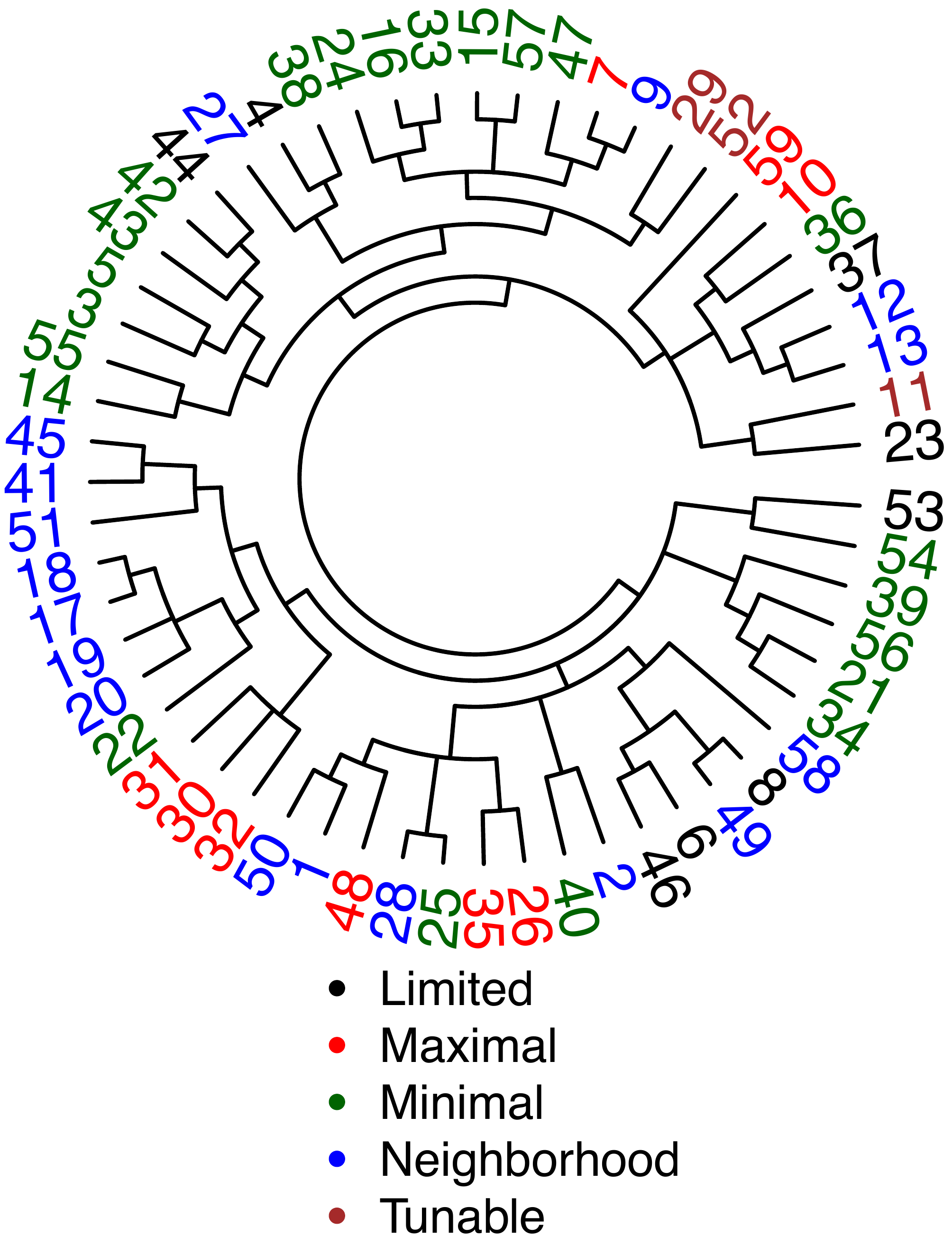}\label{fig:ka-ind}}
  \hfill
  \subfloat[Technique]{\includegraphics[width=0.224\textwidth,valign=t]{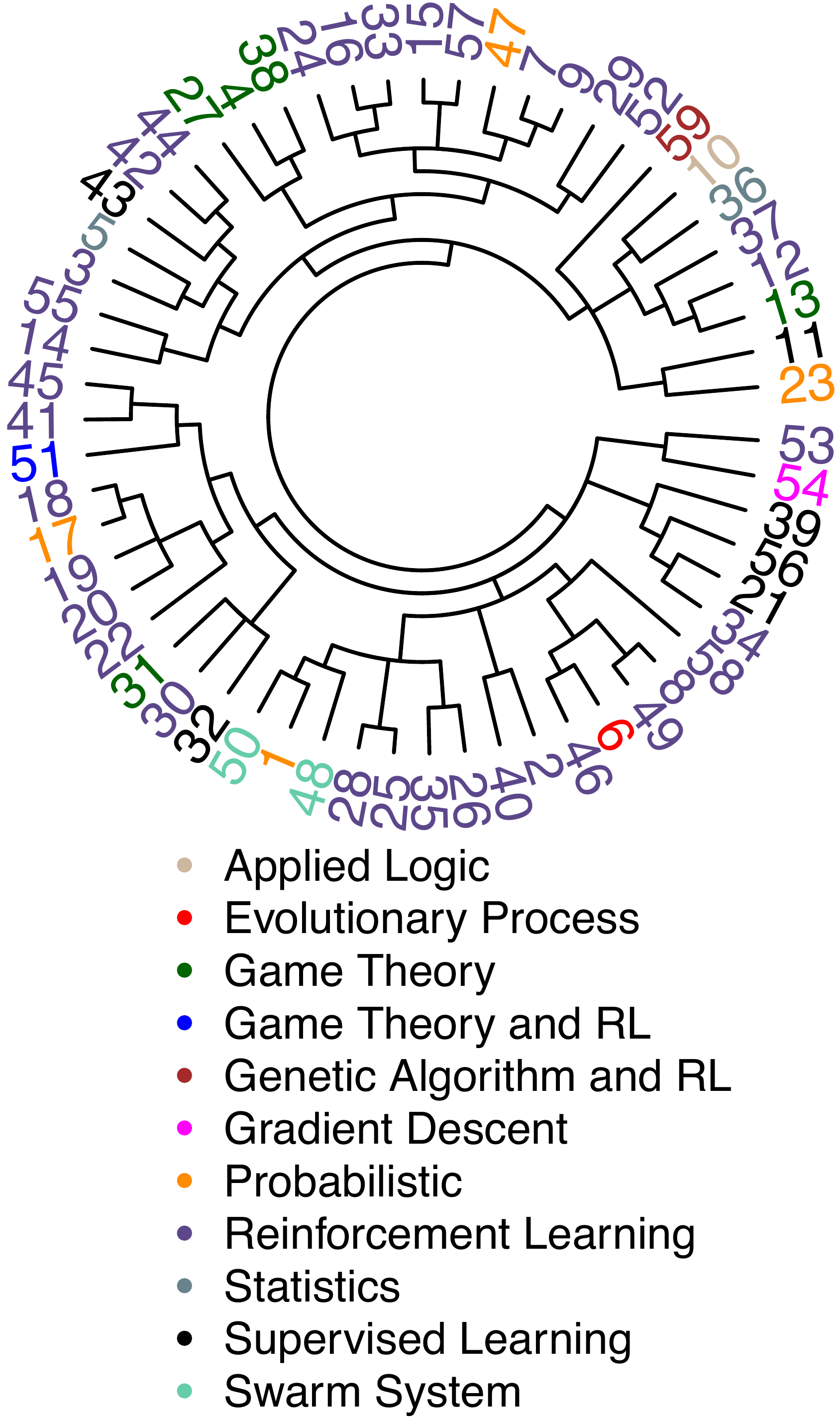}\label{fig:tec-ind}}
     \hfill
  \subfloat[Domain]{\includegraphics[width=0.224\textwidth,valign=t]{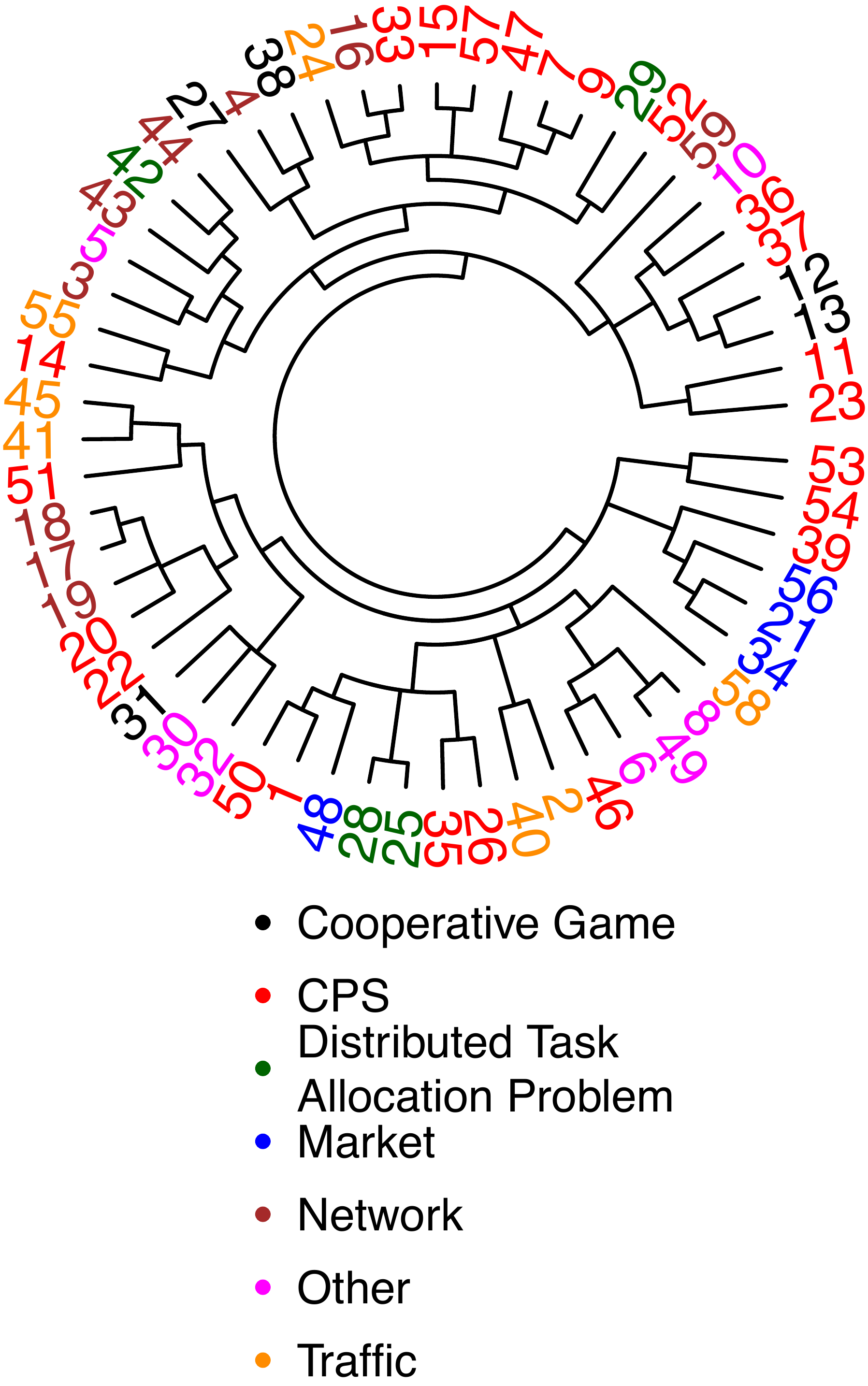}\label{fig:dom-ind}}
\caption{HAC results: evaluation (a-b), choices of $K$ (c-f), design dimensions mapped to HAC (g-o).}
\label{fig:custering_outcome}
\end{figure*}

\section{Methodology}
\label{sec:methodology}

\noindent
In this section, we present how we extended our systematic literature review on learning-enabled CAS~\cite{DAngelo:2019} and use its results as input for our data-driven methodologies. Then we explain the two employed data-driven methodologies, i.e., Hierarchical Agglomerative Clustering (HAC) and Multiple Correspondence Analysis (MCA).

Clustering allows us to capture the set of design choices applied on CAS based on past experience (i.e.,~the state of the art). The rationale is that clustering analysis can be used to detect design patterns and reduce the complexity of the design space. Correspondence analysis, on the other hand, focuses on identifying the correlation between the design dimensions. This allows capturing system constraints that impose certain design choices. 
As a result, the engineer obtains better insights about the interplay and interactions between different design dimensions. The relevance and effectiveness of these techniques for system design decisions has been shown in~\cite{Ballandies2018}.

\begin{table}[t]
\caption{DIMENSIONS OF LEARNING-BASED CAS}
\centering
\begin{tabular}{p{0.21\columnwidth}p{0.68\columnwidth}} \toprule
\textbf{Dimension} & \textbf{Description}\\ \midrule
Application \newline Domain & The domain for which a CAS is developed\\ \midrule
Autonomy & The agents' ability to act autonomously or in need of a supervised entity \\ \midrule
Knowledge \newline Access & The amount of information available to an agent from its peers or the environment  \\ \midrule
Behaviour & Agent's comportment toward self-goals and system-wide goals \\ \midrule
Emergent Behaviour & Whether the collective demonstrates a behaviour different than the one of single agents
\\ \midrule
Cooperative Agent & Whether agents are cooperating \\ \midrule
Learning \newline Technique & The technique used by agents to exhibit learning \\ \midrule
Trigger First & The initial knowledge used to instantiate learning models \\ \midrule
Trigger Update &Criteria for updating the learning models\\ \bottomrule
\end{tabular}
\label{tab:attributes}
\end{table}

\subsection{Data Acquisition}
\label{sec:data-acq}
\noindent
In~\cite{DAngelo:2019}, we conducted a systematic literature review of $52$ studies related to learning-based CAS. 
The investigated papers are classified based on their choices for the nine design dimensions envisioned for learning-enabled CAS. Table~\ref{tab:attributes} enumerates these nine design dimensions, which are used to define the design space of CAS. %and classify the reviewed papers. 
These dimensions are the result of a thorough discussion and analysis of the domain at the GI-Dagstuhl seminar on Software Engineering for Intelligent and Autonomous Systems~\cite{Gerasimou:2019}.

This paper extends the literature review of~\cite{DAngelo:2019} by following the same search and analysis methods.
The analysis of the most recent studies added 7 additional research papers. As a result, 59 studies are included in our updated systematic literature review\footnote{The updated review and replication package of the analysis can be seen at: \url{https://mi-da.github.io/learning-to-learn-CAS/}}. A vector including the nine design choices of a paper (see Table~\ref{tab:attributes}) constitutes a data point for our data-driven analysis. Accordingly, the considered dataset has $59$ data points.
%which our data-driven methodologies use as a dataset.

\subsection{Hierarchical Agglomerative Clustering (HAC)}
\label{sec:clustering}

\noindent
As a starting point, HAC~\cite{Lukasova:1979} considers each observation (i.e.,~each reviewed paper in our case) as a separate cluster. Then, HAC incrementally identifies and merges the two most similar clusters. The notion of similarity between papers refers to similarity between their design dimensions~(see~\autoref{tab:attributes}). Since the dimensions identified in our systematic literature review are categorical, we adopt the Gower Distance~\cite{Gower:1971} measure, which is a simple but widely applied distance metric suitable for categorical data. In particular, the distance $d(i,j)$ for any pair of inputs $x_i$ and $x_j$ across the examined number of dimensions $M$ is given by 
$d(i,j) = 1/M \times \sum_{m=1}^M d_{ij}^m$ where $d_{ij}^m = 0$ if $x_i^m = x_j^m$; and $d_{ij}^m = 1$ otherwise. Evaluating other distance metrics is outside the scope of this paper.

A typically applied method for identifying the desired number of final clusters in HAC (given by $K$) involves using clustering validation criteria~\cite{mullner2011modern}.
% To identify the desired number of final clusters in HAC (given by $K$) clustering validation criteria should be employed.
In our analysis, we rely on the silhouette values~\cite{Rousseeuw:1987} and the bootstrap method~\cite{Field:2007}. The results of clustering using this method are depicted in~\autoref{fig:custering_outcome}.

The silhouette value is a measure of data consistency, where higher values represent higher coherence between the data points within a cluster. Figure~\ref{fig:sil} shows the silhouette values for HAC up to 10 clusters. The plot suggests that after five clusters, the consistency of data points within each cluster drops and reaches a local maximum with $K=9$. 

The bootstrap analysis enables the assessment of the stability of the considered number of clusters by investigating how easily clusters dissolve.
To conduct the bootstrap analysis, we select the clusters with average silhouette width greater than or equal to $0.18$, i.e.,~the average silhouette width with $K=9$ (see the dotted line in Figure~\ref{fig:sil}). Lower silhouette values result in weak or artificial structures.
The results of the bootstrapping are presented in Figure~\ref{tab:bootstrap}. For each number of considered clusters $K$ we report: the vector of cluster stabilities (values close to $1$ indicate stable clusters) and the number of times each cluster is dissolved after $100$ re-sampling~(clusters that are dissolved often are unstable). The results suggest that $K=2$ forms two stable clusters that are never dissolved. $K=3$ and $K=4$ introduce mild degrees of instability, while $K=5$ and $K=9$ result in high instability. 

Figures~\ref{fig:2cl}--\ref{fig:9cl} show how the reviewed papers, distinguished by their id, are partitioned in 2, 3, 4, and 9 clusters, respectively. In particular, each figure shows how cluster partitions, identified with different colors, map to the dendrogram tree generated by the HAC. %For instance in~Figure~\ref{fig:2cl} two clusters are identified: the top cluster ranges from paper with id 14 to paper with id 23, the bottom cluster, from paper with id 45 to paper with id 53.
For instance,~Figure~\ref{fig:2cl} depicts the result of clustering for $K=2$, where the two clusters are identified with blue~($31$ papers) and red~($28$ papers).

Figures~\ref{fig:autonomy-ind}--\ref{fig:dom-ind} illustrate the results of clustering the dataset based on each design dimension introduced in Table~\ref{tab:attributes}. %how the \MD{nine design dimensions map to the dendrogram}. 
In particular, each figure shows how the concrete values of a specific design dimension partition the papers in the dataset. Different colors in each dendrogram represent a value for the considered design dimension.
%
%identified with different colors in the legend, characterize the papers arranged in the tree. 
%\MD{As an example, referring to Figure~\ref{fig:coop-ind}},
In~Figure~\ref{fig:coop-ind}, for instance, the dendrogram partitions the papers according to the design dimension Cooperative Agent. The paper ids (i.e.,~data points) in red employ cooperative agents while the paper ids in black exclude cooperation among the employed agents.
%with id 14 to 23 (top arc of the circle) and 40 are characterized by non cooperation between agents, the others employ cooperative agents. 

The partitions identified by the design dimension Cooperative Agent in~Figure~\ref{fig:coop-ind} are similar to the those depicted by the clustering analysis with $K=2$ in Figure~\ref{fig:2cl} (except for paper $40$).
%
%\MD{In a similar fashion, by looking at how design dimensions relates to cluster partitions, we are able to derive design guidelines and capture the relevant set of design choices applied on CAS based on past experience.}
Similarly, mapping the results of the clustering based on the design dimensions and the results of HAC can be leveraged to capture the relevant set of design choices for CAS based on past experience.

%The result of such analysis is shown in Section~\ref{sec:application}.
%\TODO{I don't understand what is the point of this last paragraph here? Also we should avoid forward referencing. }

\subsection{Multiple Correspondence Analysis (MCA)}
\label{sec:mca}

\begin{figure}[t]
%\vspace{-0.10cm}
\centering
\includegraphics[width=0.85\columnwidth]{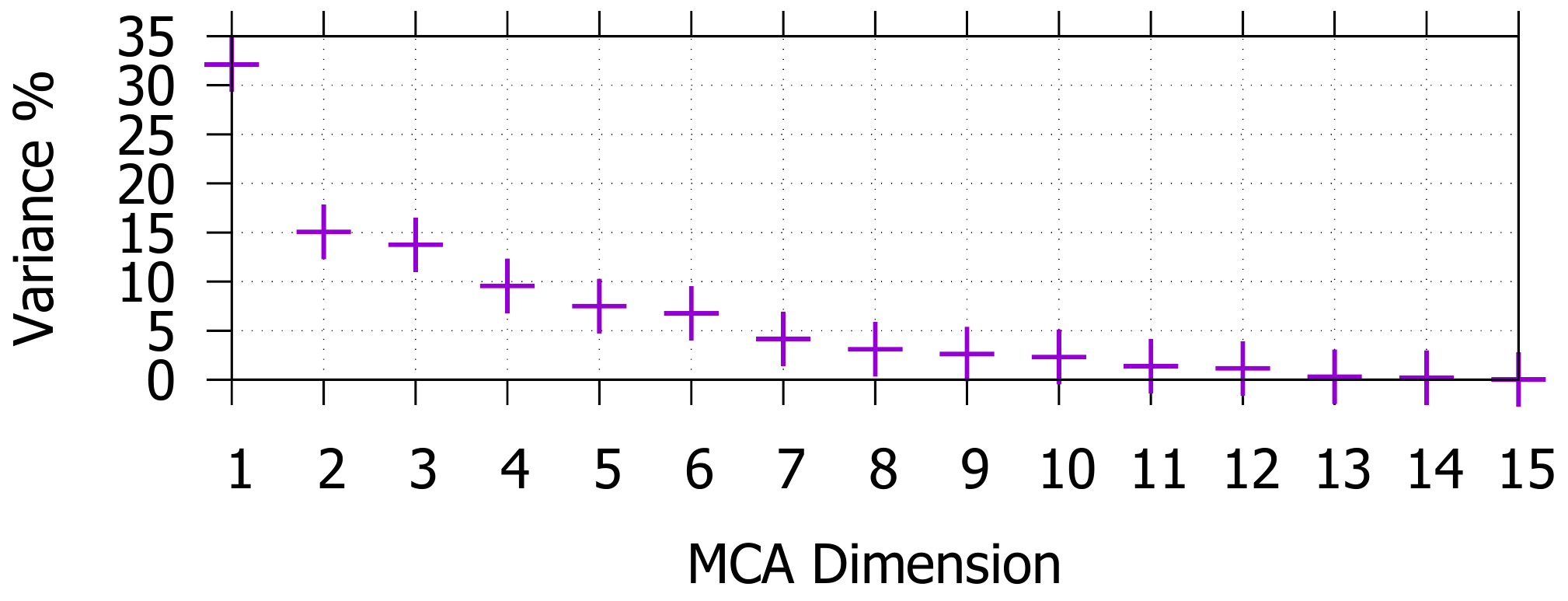}
\caption{Variance retained by MCA dimensions.}
\label{fig:mca-results}
\end{figure}

\noindent
MCA is a generalization of the principal component analysis~(PCA) for categorical data, which aims to summarize the underlying structures in the fewest possible dimensions~\cite{Greenacre:2006}. In particular, MCA identifies new latent dimensions, which are a combination of the original dimensions and hence can explain information that is not directly observable.

Figure~\ref{fig:mca-results} depicts the variance of the new dimensions (i.e., principal components) identified by MCA after applying the optimistic Benzécri correction~\cite{Greenacre:2006}. The larger variance of the dimensions indicates capturing more meaningful correlations by the considered dimensions.

\begin{figure}[t]
%    \vspace{-0.30cm}
    \centering
    \hspace{-0.43cm} 
    \subfloat[MCA Dimension 1]{\includegraphics[width=0.5\columnwidth]{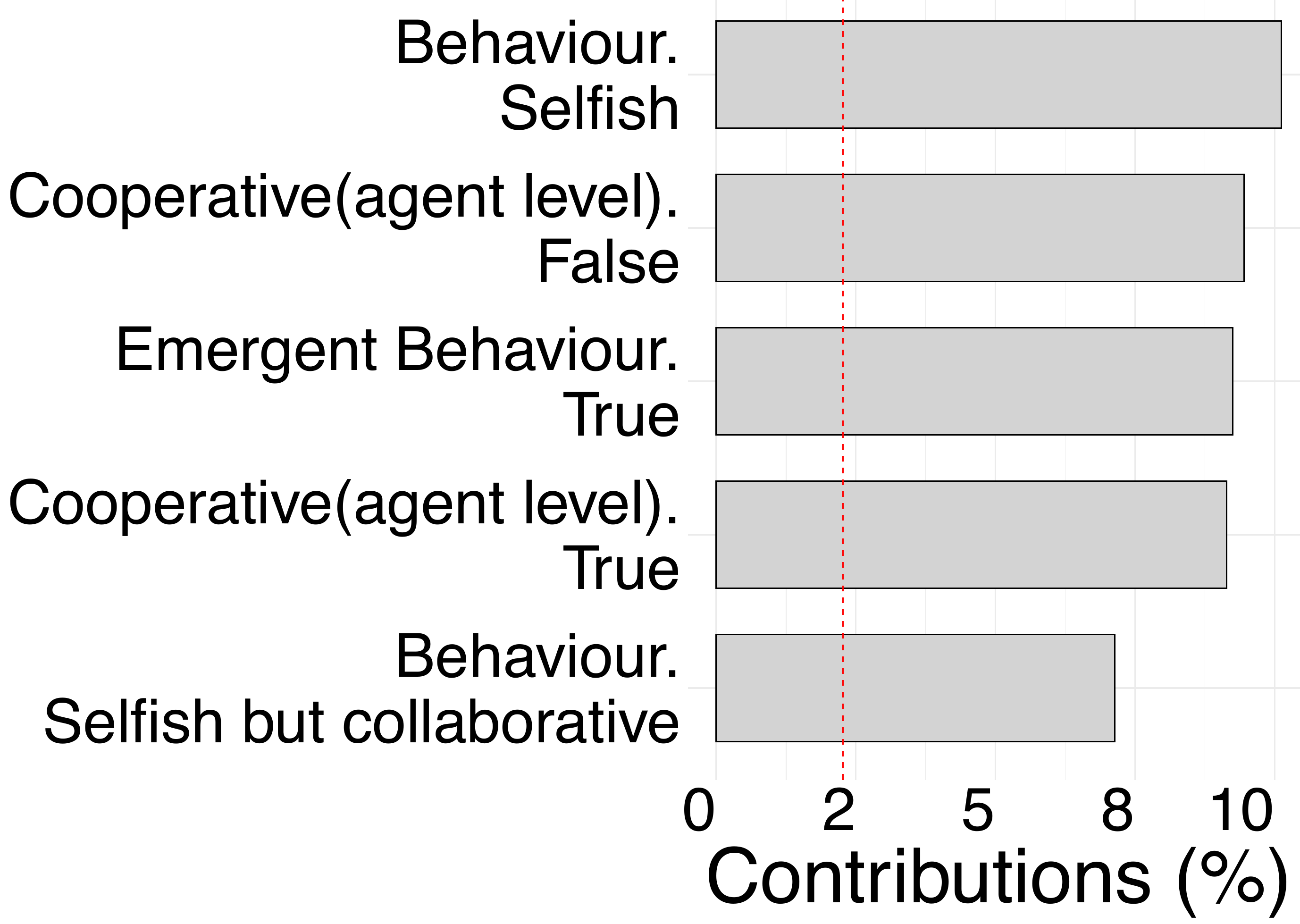}\label{fig:pc1}}%
    \subfloat[MCA Dimension 2]{\includegraphics[width=0.5\columnwidth]{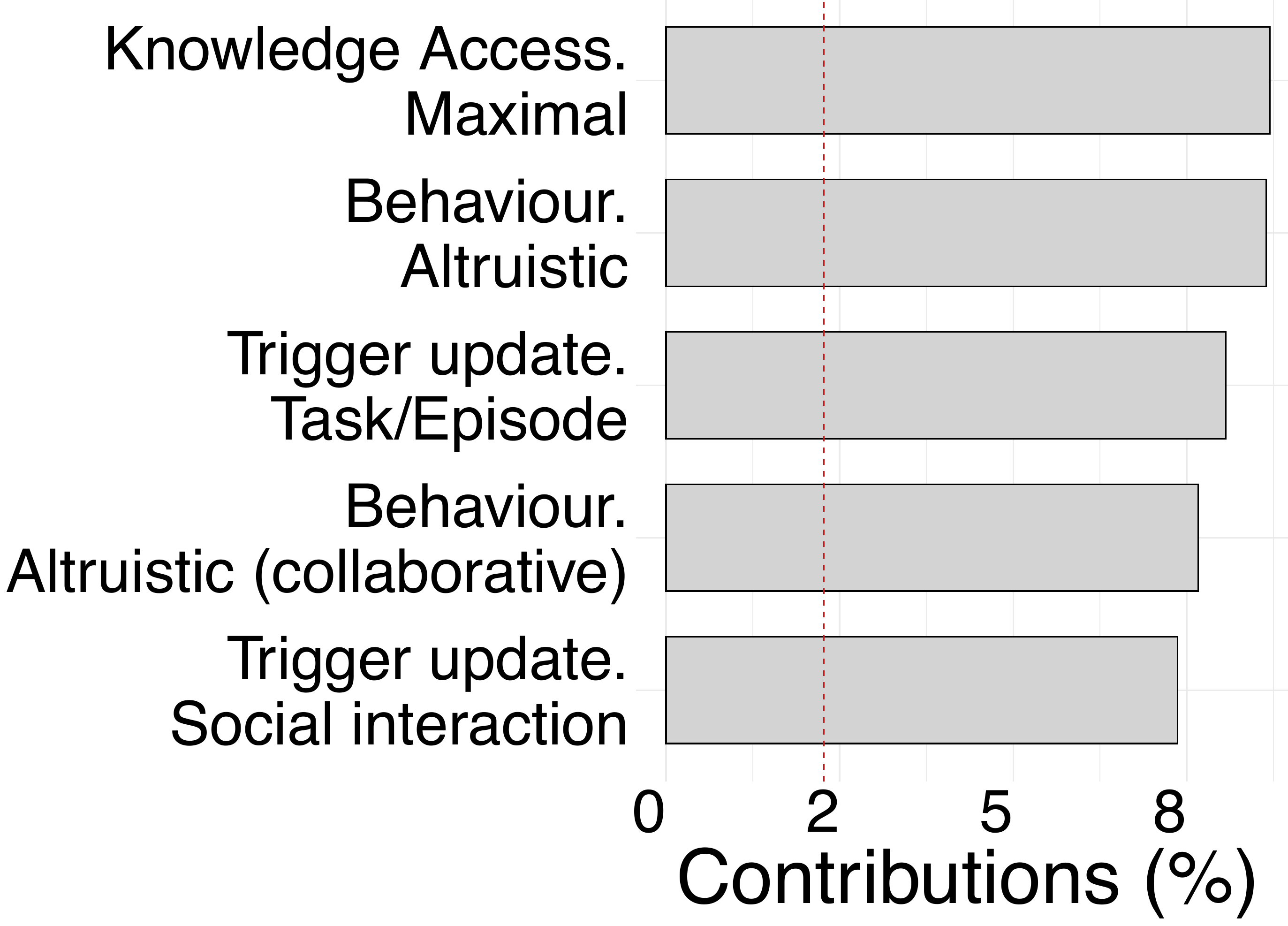}\label{fig:pc2}}%
    
   \hspace{-0.43cm} 
   \subfloat[MCA Dimension 3]{\includegraphics[width=0.515\columnwidth]{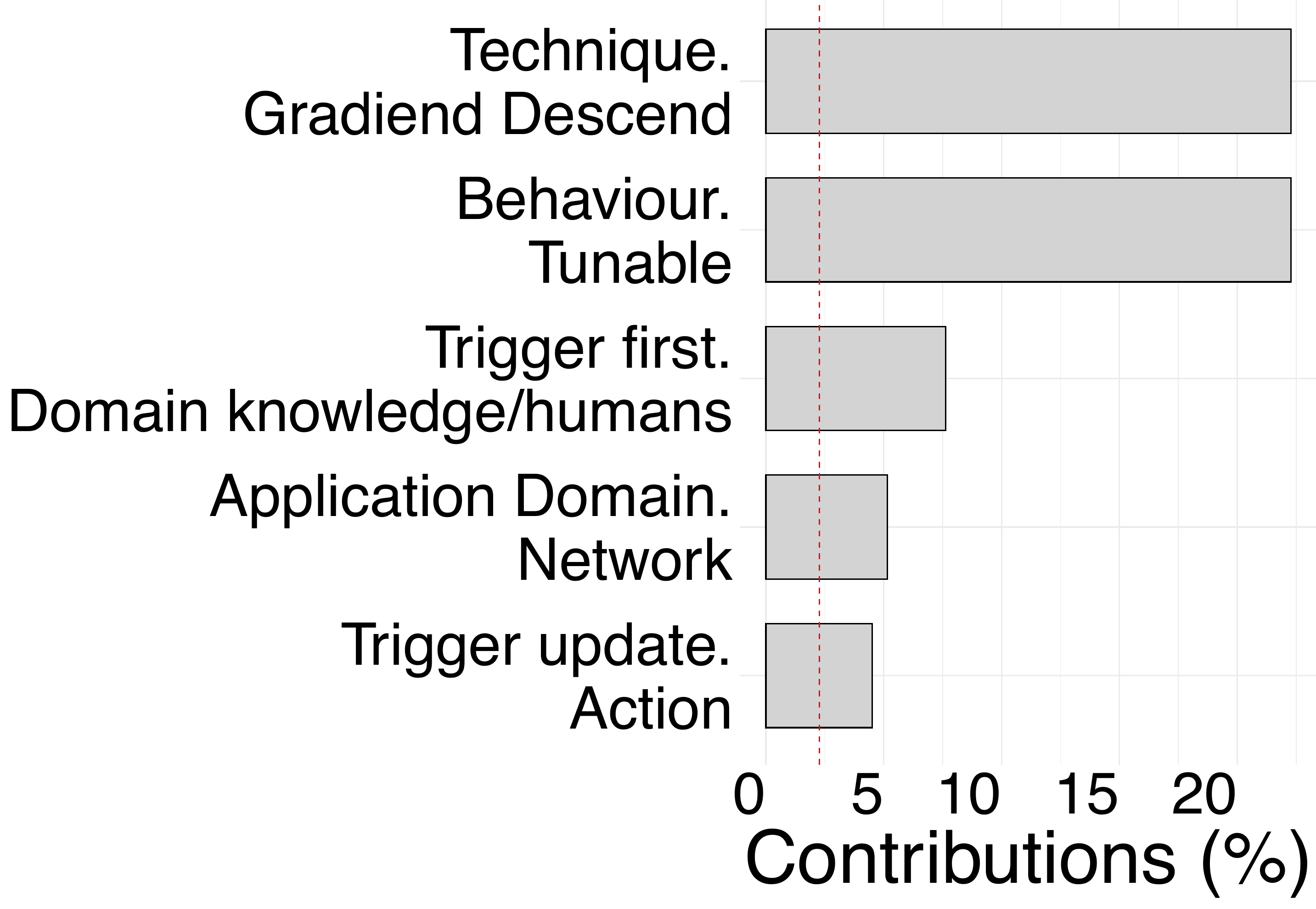}\label{fig:pc3}}%
   \subfloat[MCA Dimension 4]{\includegraphics[width=0.515\columnwidth]{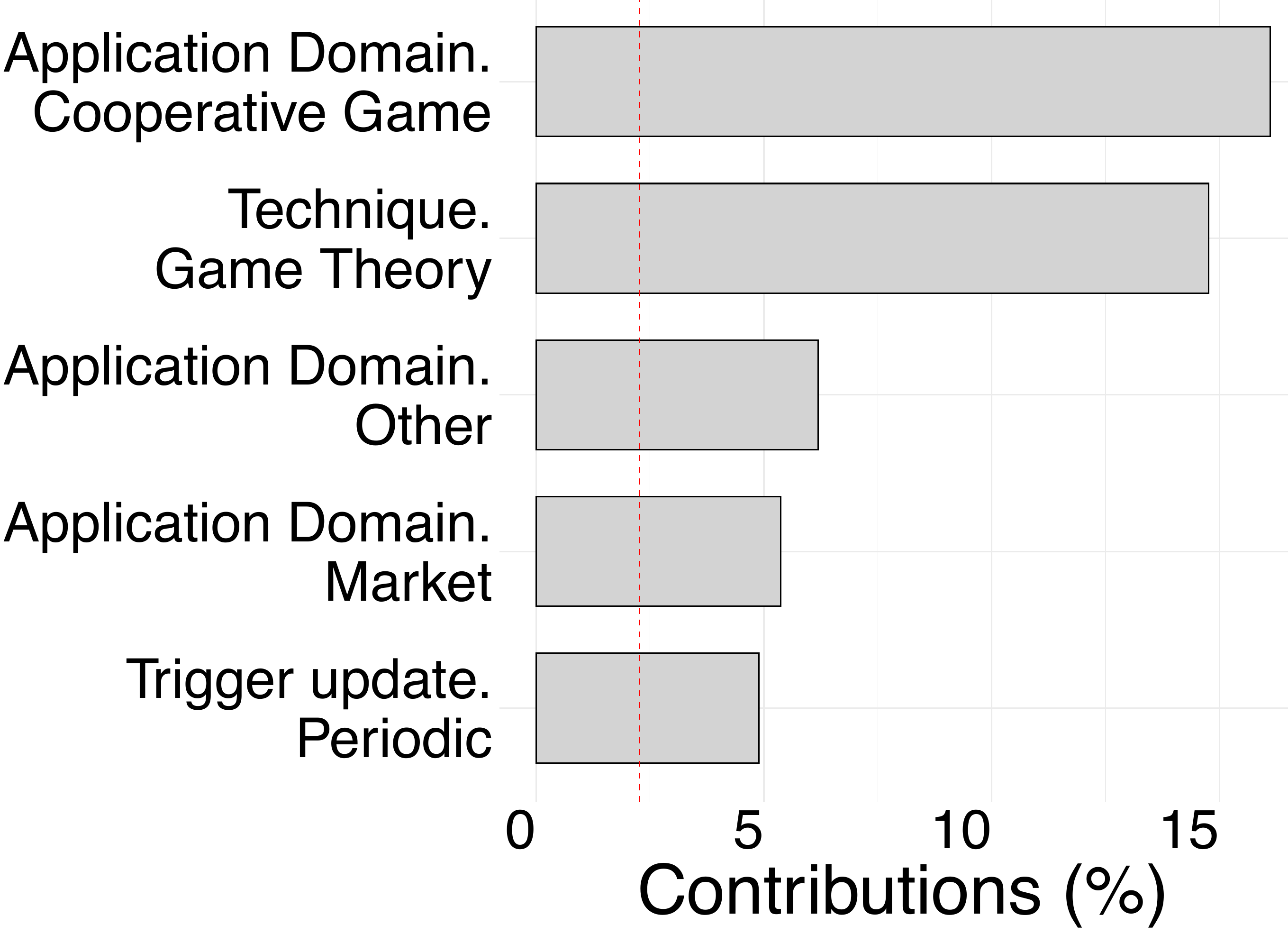}\label{fig:pc4}}% 
   \caption{Contribution of design dimensions to MCA dimensions.}%
   \label{fig:mca-pc}
\end{figure}

In order to have a good intuition of the MCA results, it is necessary to choose the number of components to retain and observe how the design dimensions of CAS map to the new identified dimensions. Following the average rule introduced by Lorenzo-Seva et. al.~\cite{Lorenzo-Seva:2011}, we kept all the dimensions with variance greater than $7\%$ (i.e., four dimensions are retained).
Figure~\ref{fig:mca-pc} depicts the contributions (in percentage) of the design dimensions' values to the definition of the MCA dimensions. Figures~\ref{fig:pc1}--\ref{fig:pc4} show only the five most contributing variables, as small contributions imply low relevance. The dashed lines represent the expected contribution if all the values of the design dimensions would contribute equally to the definition of the MCA dimension.

\section{Application}
\label{sec:application}
\noindent
We first present the observations and guidelines emerging from applying our data-driven analysis. Then, we present the extracted reasoning knowledge in the form of a decision tree that can be used both as a recommender system and as a mechanism to identify design gaps.
\begin{figure}[b]
\includegraphics[width=0.9\columnwidth]{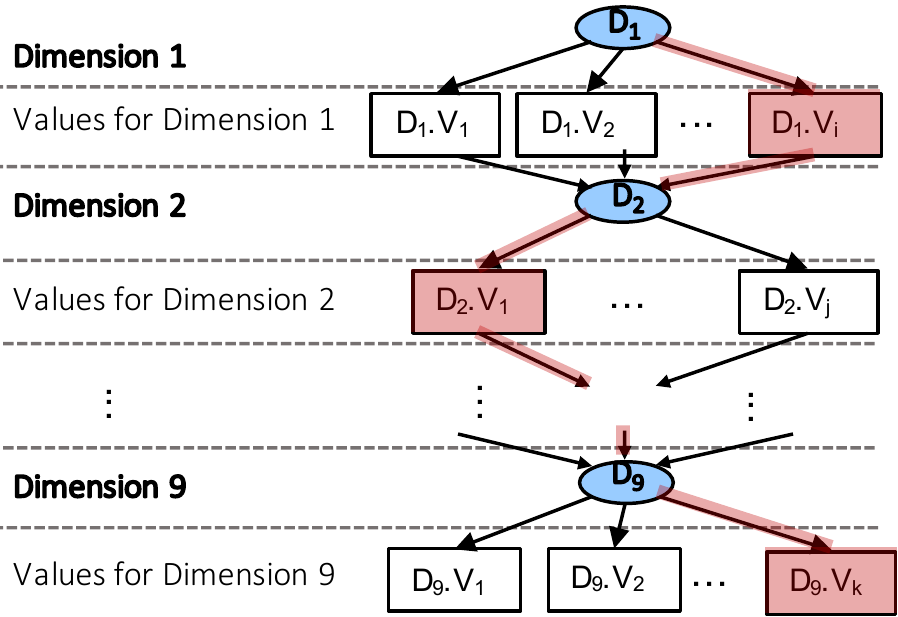}
\centering
%\caption{Decision Tree Path on Design Dimensions' values}
\caption{Decision tree representation of reasoning knowledge.}
\label{fig:treePrototype}
\end{figure}

\begin{figure}[t]
\includegraphics[width=1\columnwidth]{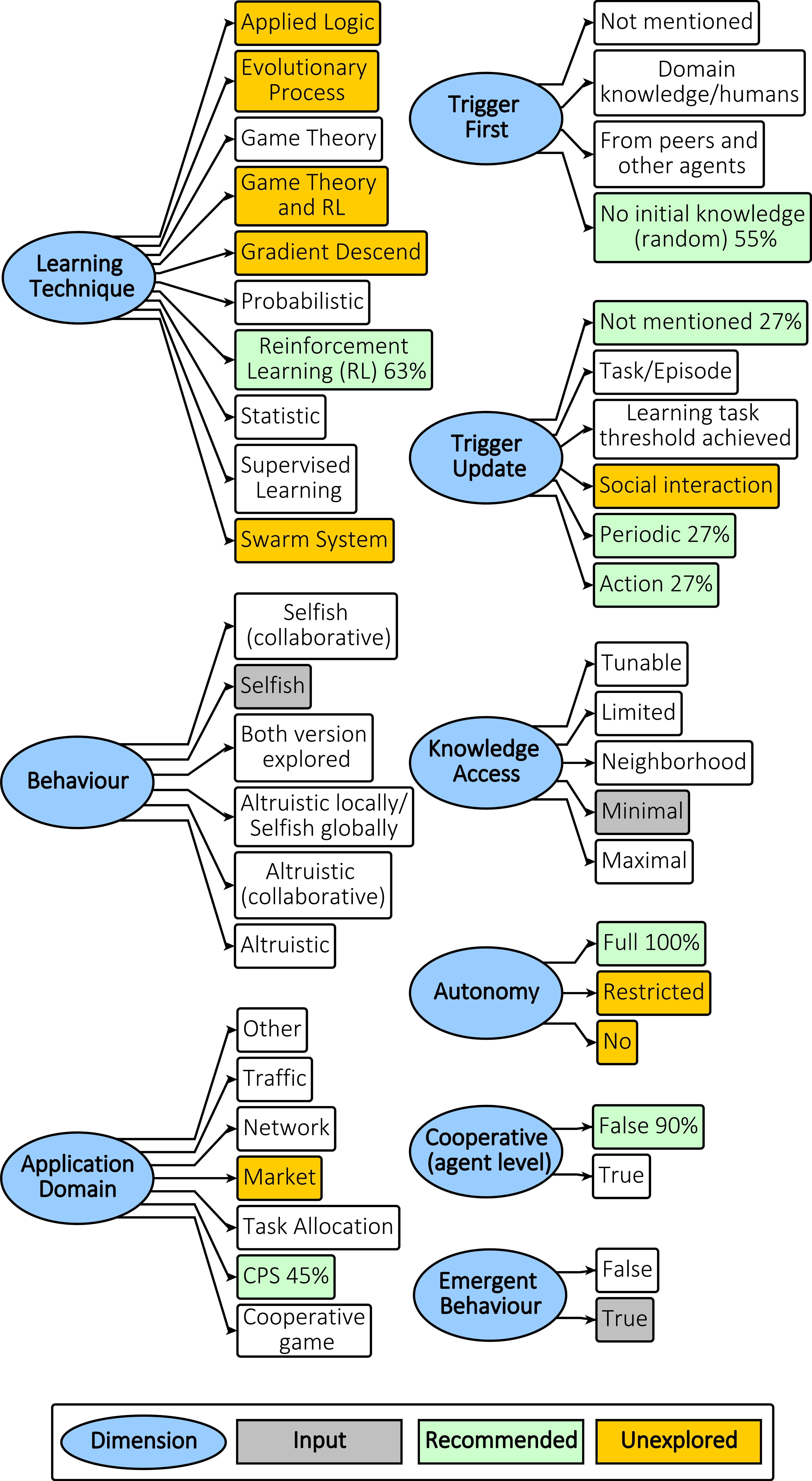}
\centering
\caption{Design dimensions as decision nodes.}% and their constituting values }
\label{fig:D-Tree}
\end{figure} 
\subsection{Observations and Guidelines}
\noindent
Emergent behaviour and cooperative agents are two of the most important design dimensions identified by MCA (see Figure~\ref{fig:pc1}) that promote the formation of two stable clusters (see Figure~\ref{fig:2cl}). 

Our analysis highlights that designing non-cooperative (see Figure~\ref{fig:coop-ind})  and selfish agents (see Figure~\ref{fig:self-ind}) often results in a collective emergent behaviour (see Figure~\ref{fig:eb-ind}).

In contrast, when interactions are introduced, it is often the case that designers have a target system behavior in mind. This is derived from the observation that systems using altruistic approaches or adopting a certain level of collaboration are characterized by employing cooperative agents and no emergent behaviour.

MCA finds correlation between the two dimensions indicating the triggers for instantiating~(i.e.,~trigger first) and refining the learning models~(i.e.,~trigger update). The process employed for learning a new or updating the existing model is a major factor affecting the ability of agents within CAS to operate in uncertain environments.

Our 
%clustering
analysis suggests that in the presence of maximum knowledge access (see Figure~\ref{fig:ka-ind}), domain or human knowledge is often used to instantiate the learning model and the refinement is typically through episodes or task-based (see Figures~\ref{fig:first-ind}--\ref{fig:update-ind} and Figure~\ref{fig:pc2}).
On the other hand, we identify a group of systems that are characterized by minimal or neighborhood knowledge access and, as a result, their learning models operate under high uncertainty. In this case, the learning model is typically instantiated without any prior knowledge and the refinement is threshold-based or via social interactions. Our analysis highlights how agents often rely on their peers' knowledge in the presence of learning uncertainty to construct a bigger picture of the environment as a means to tackle uncertainty. 

We summarize our findings by proposing the following design guidelines, which are the result of how existing CAS are usually engineered.

\textbf{Design for System-wide vs.\ Agent-level Goals.} If the objective of the CAS is to fulfill a system-wide goal, then: \emph{(i)}~cooperative agents shall be used; \emph{(ii)}~agents should exhibit altruistic or collaborative behaviour; and \emph{(iii)}~there should be a degree of knowledge exchange among agents (i.e.,~minimal level of knowledge access should be avoided). The engineer should decide/reason about the trade-off between the desired level of knowledge and the introduced cost in terms of performance. 

In contrast, designing a CAS with agent-level objectives in mind: \emph{(i)}~eliminates the need for employing cooperative agents; \emph{(ii)}~agents demonstrating selfish behaviour (hence, prioritizing agent-level goals) should be employed; and \emph{(iii)}~knowledge access among agents is not required since no coordination is needed. In such a scenario, the engineer can expect the (implicit) system-wide goal to be fulfilled as an emergent behaviour. 

\textbf{Access to Training Data/~Domain Knowledge.} The choice of a learning technique can be greatly affected by the availability of training data or domain expert knowledge for model instantiation and update. 
When access to sufficient data for model-based learning techniques is limited, model-free techniques should be employed. 
When no sufficient data is available, the refinement should be threshold-based or via social interactions. Domain expert knowledge can be leveraged to set the model refinements trigger to episodic or task/action-based. Finally, the choice of the learning technique should be independent of the application domain of interest as we observe no correlation between these two design dimensions.

\subsection{Mining the Knowledge}
\noindent
We employ decision tree modeling~\cite{APTE1997197} to generate a decision tree from the collected data. The identified design dimensions introduced in Table~\ref{tab:attributes} form the nine decision nodes of the tree as depicted in Figure~\ref{fig:treePrototype}. Each dimension $D_i$ is further decomposed into multiple values $D_i.V_j$. We collected the values for each design dimension from our survey in~\cite{DAngelo:2019}. Figure~\ref{fig:D-Tree} expands the design dimensions introduced in Table~\ref{tab:attributes} to their different values.

Representing the reasoning knowledge as a decision tree provides a top-down scheme to explore the design space of learning-based CAS for the CAS designers. The decision tree can be traversed 
%by
starting from the root node (i.e.,~\textbf{$D_1$} in~Figure~\ref{fig:treePrototype}) and making decisions among the available choices for each dimension \textbf{$D_i$}, (i.e.,~$D_i.V_j$ in~Figure~\ref{fig:treePrototype}) until all the dimensions are visited. The highlighted trajectory in~Figure~\ref{fig:treePrototype} shows an exemplary path in the decision tree where value $V_i$ is selected for dimension \textbf{$D_1$}, value $V_1$ is chosen for \textbf{$D_2$}, and for dimension \textbf{$D_9$} value $V_k$ is selected. The path can be summarized as $(D_1.V_i , D_2.V_1 , ... , D_9.V_k)$.

%Figure~\ref{fig:D-Tree} expands the design dimensions introduced in Table~\ref{tab:attributes} to their different values which are also captured by our survey in~\cite{DAngelo:2019}.
Each dimension \textbf{$D_1$}--\textbf{$D_9$} in~Figure~\ref{fig:treePrototype} is mapped to a design dimension introduced in Table~\ref{tab:attributes} and is depicted in~Figure~\ref{fig:D-Tree} as a blue decision node. Each decision node in~Figure~\ref{fig:D-Tree} is decomposed into its constituting values. %represents a decision node of the decision tree whose children are the different values. 

We elaborate on the application of the reasoning knowledge as a decision tree by presenting a CAS design scenario.
To this end, consider a design scenario where an engineer of a learning-based CAS wants to minimize the data exchange for security reasons. As a result, the engineer's goal is to design a CAS where agents have minimal knowledge access. To support this choice, the agents preferably need to have selfish behaviour, since they are not exchanging any information and have no global view. Consequently, the engineer expects to observe some emergent behaviour, resulting from the collective actions of the agents. The engineer's requirements serve as input to the decision tree presented in Figure~\ref{fig:treePrototype}. The designer's input provides the values for three out of nine dimensions. This input sequence can be represented in $D_i.V_j$ format and formalized as: ($Behaviour.Selfish$, $Knowledge Access.Minimal$, $Emergent Behaviour.True$), shown as gray nodes in~Figure~\ref{fig:D-Tree}. %, representing our collected reasoning knowledge. 
 Consequently, taking the input sequence into account, we can %suggest the values for the remaining  \SG{six} dimensions with a confidence value (based on similar past experiences) and also depict the unexplored design combinations in the following manners: 
 navigate the rest of the decision tree (i.e.,~the remaining six dimensions) in the following manners.

\textbf{Decision Tree as Recommender System.} Based on the provided input, we can query similar past experiences, that is, data items with the same values for the three specified dimensions. As a result, we can  suggest the values for the remaining six dimensions with a confidence metric based on similar past experiences. The confidence metric for each value represents the percentages of the similar past experiences employing the same value.% \todo{there is some repetition of "similar"}

%(gray nodes in~Figure~\ref{fig:D-Tree}),
The green nodes in~Figure~\ref{fig:D-Tree} depict the most common choices for the values of each dimension among the past experiences as recommendation for the provided input. The decision tree suggests %~(green nodes in~Figure~\ref{fig:D-Tree})
to the engineer a fully autonomous system with no cooperation among the agents. Reinforcement learning is a good candidate technique to adopt, as it is used in $63\%$ of the studies engineering similar systems. The initial learning trigger should be performed randomly if no initial knowledge is available or from other agents. There are no outstanding suggestions on how to update the learning model. Similar systems are used in the CPS domain. 
    
\textbf{Spotting the Gaps via Decision Tree.} The decision tree can also be navigated to spot unexplored combinations of dimensions~(see the yellow nodes in~Figure~\ref{fig:D-Tree}), which might result in novel design decisions. For instance, for the provided input in the envisioned scenario, social interactions are never used to update the learning model in the identified scenario. Learning techniques such as applied logic and gradient descend also have never been applied in combination with the given input. 

The application of the decision tree is not limited to the existence of an input sequence. In cases where the engineer has no preferences for any of the design dimensions, the tree can be used either as a recommender or to spot the gaps for all nine dimensions in a similar manner as described above. For the case of the recommender system, the confidence metrics are calculated among all the data points.

\section{Conclusion and Future work}
\label{sec:conclusion}
\noindent
Using the data acquired from a systematic literature review of relevant research on learning-based CAS, we leverage on data-driven methodologies to reason about design principles of CAS. We present the data in the form of a decision tree that can be used to mine existing patterns or explore possible novel design choices.

The techniques used in this paper can be applied to an extended dataset, including more dimensions.
Building an interactive web-based system that can be further enriched by researchers and provide automated analysis such as those presented in this paper is a subject of our future work. Moreover, the extension of our analysis to capture insights about CAS evolution across communities and over the course of time is another planned future work.

\bibliographystyle{IEEEtran}
\bibliography{biblio}

\end{document}